**Electronic structure of pristine and K-doped solid picene: Non-rigid-band change and its implication for electron-intramolecular-vibration interaction**

H. Okazaki,[1,2] T. Wakita,[1,2] T. Muro,[3] Y. Kaji,[1] X. Lee,[1] H. Mitamura,[1] N. Kawasaki,[1] Y. Kubozono,[1] Y. Yamanari,[1] T. Kambe,[1] T. Kato,[4] M. Hirai,[1] Y. Muraoka,[1,2] and T. Yokoya[1,2]

[1]*The Graduate School of Natural Science and Technology, Okayama University, 3-1-1 Tsushima-naka, Okayama 700-8530, Japan*

[2]*CREST, Japan Science and Technology Corporation (JST), 3-1-1 Tsushima-naka, Okayama 700-8530, Japan*

[3]*Japan Synchrotron Radiation Research Institute (JASRI) /SPring-8, 1-1-1 Kouto, Sayo, Hyogo 679-5198, Japan*

[4]*Institute for Innovative Science and Technology, Graduate School of Engineering, Nagasaki Institute of Applied Science, 3-1, Shuku-machi, Nagasaki 851-0121, Japan*

**Abstract**

We use photoemission spectroscopy to study electronic structures of pristine and K-doped solid picene. The valence band spectrum of pristine picene consists of three main features with no state at the Fermi level ($E_F$), while that of K-doped picene has three structures similar to those of pristine picene with new states near $E_F$, consistent with the semiconductor-metal transition. The K-induced change cannot be explained with a simple rigid-band model of pristine picene, but can be interpreted by molecular orbital calculations considering electron-intramolecular-vibration interaction. Excellent agreement of the K-doped spectrum with the calculations points to importance of electron-intramolecular-vibration interaction in K-doped picene.

## 1. Introduction

Carbon systems have attracted much attention because of the discovery of new superconductors with relatively high superconducting transition temperatures; heavily boron-doped diamond with $T_c$ < 10 K [1,2], graphite intercalation compound (GIC) superconductor $CaC_6$ with $T_c$ = 11.5 K [3], and alkali-doped fullerene $Cs_3C_{60}$ with $T_c$ = 38 K under 7 GPa [4]. Recently, a new type of superconductors was discovered in a solid of aromatic molecule picene, with $T_c$ = 7 and 18 K when doped with potassium (K) [5]. This is the first discovery of superconductivity in aromatic molecular crystals and, thus, motivates researchers to search for new aromatic molecular superconductors with higher $T_c$ as well as to understand the mechanism of the superconductivity. A picene molecule consists of five benzene rings sharing some of their edges, forming a zigzag structure, and solid picene has a layer structure stack to *c* axis, where each layer has a herringbone structure with the picene molecules inclining a little from the *ab* plane [6]. Thus, the structure is two dimensional. Solid picene is remarkable for its physical properties; wide band gap of 3.3 eV [7] and high carrier mobility of greater than 3 $cm^2$ $V^{-1}$ $s^{-1}$ by exposure to $O_2$ [8]. However, neither the electronic structure of the pristine nor doped picene has been reported, yet. It is, therefore, crucial to study experimental electronic structure of pristine picene and its evolution with doping in order to understand the mechanism of superconductivity.

For alkali-doped superconducting fullerene, which is similar to molecular crystal as picene, it is considered that the electrons of doped alkali atoms are transferred to the threefold degenerated lowest unoccupied molecular orbital (LUMO) bands of fullerene. However, the electrons do not rigidly occupy the LUMO bands of fullerene, instead the LUMO bands split into the occupied and unoccupied states [9] by Jahn-Teller distortion [10], which originates from a coupling between electrons doped into the LUMO and intramolecular phonon of a fullerene molecule. The electron-intramolecular-vibration coupling is considered to be essential in understanding the physical properties of fullerides including superconductivity [10]. Therefore, the role of electron-intramolecular-vibration interaction in K-doped picene should be investigated in order to understand the superconductivity.

In this paper, we report photoemission spectroscopy (PES) of pristine and K-doped picene films in order to elucidate the evolution of electronic structure by K doping. A valence band spectrum of K-doped picene shows new states near $E_F$ besides several structures that can be related to those of pristine picene. The states near $E_F$ in K-doped picene can be ascribed to states derived from the LUMO of pristine picene but is not interpreted by a simple rigid shift of the LUMO of pristine picene. The valence band spectrum of K-doped picene is in excellent agreement with the molecular orbital calculation of negatively charged picene including electron-intramolecular-vibration interactions. This result indicates importance of electron-intramolecular-vibration interactions in K-doped picene, which can play a crucial role for the metallic properties leading to superconductivity.

2. Experimental

Picene was prepared by photosensitization of 1,2-di(1-naphthyl)ethane with 9-fluorenone in chloroform. The details of the procedure are described in a literature [7]. Pristine picene thin films for the measurements at Hiroshima Synchrotron Radiation Center (HiSOR) were prepared by *in-situ* deposition of picene powder on Au-coated stainless substrates under ultrahigh vacuum. X-ray diffraction of the solid pristine picene films showed that they have the same crystal structure as that of solid picene. K-doping was achieved by successive deposition of potassium onto the pristine picene films for 40 mins. The K concentration $x$ of K-doped picene ($K_xC_{22}H_{14}$) films in the region measured with present PES was estimated to be 1.0 ± 0.3 from the spectral weight ratio between K 3$p$ and C 2$s$-$p$ derived valence band measured with 100 eV, taking the ionization cross sections of K 3$p$ and C 2$p$ or C 2$s$ into consideration [11] (see later for the detail). For the measurements at SPring-8, *ex-situ* pristine picene films prepared with the same procedure as described above were used. Absence of oxygen-related signals in the pristine picene films ensured no oxygen-related adsorption.

We performed the PES measurements at two different sites in order to distinguish C 2*s* and 2*p* orbitals utilizing the difference of cross-section ratio. The PES measurement at BL-5 (Okayama University Beamline) of HiSOR was performed using photon energy of 100 eV. We used the 100 eV photon energy in order to obtain high signal to noise ratio, as the intensity of 100 eV photon energy is the highest in this beamline. Another PES measurement at BL27SU of SPring-8 was done using photon energy of 1100 eV. The energy resolutions of PES at HiSOR and SPring-8 were set to be 250 and 300 meV, respectively. All the measurements were performed at room temperature and under the base pressure better than $3 \times 10^{-8}$ Pa. $E_F$ of the samples was referenced to that of a Au film which was measured frequently during the experiments.

Molecular structures of the neutral and monoanionic picenes were optimized by using the hybrid Hartree–Fock (HF)/density-functional-theory (DFT) method of Becke [12] and Lee, Yang, and Parr [13] (B3LYP), and the 6-31G* basis set [14]. The Gaussian 03 program package [15] was used for our theoretical analyses. This level of theory is, in our experience, sufficient for reasonable descriptions of the geometric and vibrational features of hydrocarbons.

## 3. Results and discussion

Figure 1 shows valence band spectra of solid pristine picene measured using incident photon energies of 1100 eV and 100 eV, together with a calculated valence band of pristine picene. The experimental spectrum of pristine picene measured at 1100 eV consists of mainly three regions (I, II, and III): a broad and higher-intensity structure from 16 eV to 23 eV, a peak at 14.5 eV, and a lower intensity region with several structures between the Fermi level ($E_F$) and 13 eV binding energy. The intensity close to $E_F$ is negligible, consistent with the semiconducting nature of solid pristine picene. Change of electronic structure regarding semiconductor to metal transition will be discussed in relation to K-doped picene data later. It is known that valence band structures of carbon systems (graphite, diamond, and hydrocarbons [16]) are formed from C 2$s$ and 2$p$ orbitals, with the energy position of C 2$p$ closer to $E_F$. Hybridization between these orbitals makes the band width wider with unchanged relative energy, leading to the C 2$s$ dominant band at higher binding energy and the C 2$p$ dominant band closer to $E_F$, with the 2$s$-$p$ states at the region where 2$s$ and 2$p$ states overlap [16]. Thus one may tentatively assign the states at the regions I, II, and III to C 2$s$, C 2$s$-$p$, and C 2$p$ dominant states, respectively. This assignment can be checked experimentally with photon-energy dependent of valence band spectra. In the valence band spectrum at 100 eV as compared with that at 1100 eV, the intensity of the region III is strongly enhanced while that of the region I is reduced, if the spectra are normalized to the intensity of the peak at 13 eV. As the cross-section ratio of C 2$s$ electrons to C 2$p$ electrons is the order of $10^2$ for 1100 eV but is the order of unity for 100 eV [11], this agrees well with the considerations above, leading to a conclusion that C 2$s$-states dominate the density of states (DOS) from 16 to 23 eV (region I), C 2$p$-states dominate DOS from 2 to 13 eV (region III), and the peak at 14.5 eV consists of C 2$s$ states mixed with C 2$p$ states (region II).

As for the region III, one sees several fine structures. We compare the valence band spectra of pristine picene with the molecular orbital calculation of pristine picene. The vertical bars at bottom are the calculated energy levels. The symmetries of the LUMO and the highest occupied molecular orbitals (HOMO) are $a_2$ and $b_1$, respectively. The calculated spectrum is obtained from calculated molecular orbital levels convoluted with Gaussian (FWHM of 0.7 eV) to reproduce the observed spectrum. The use of a larger FWHM of Gaussian than the energy resolution suggests that energy levels of molecular orbitals are dispersed by crystallizing. The cross-sections of C 2*p* and 2*s* are not taken into consideration, assuming that they are comparable at photon energy of 100 eV. As indicated with the vertical solid lines, the observed peak structures can be corresponded to calculated structures very well, except for the sharp peak at 11 eV [17]. This result, as is expected for the molecular crystal, indicates that the valence band of solid pristine picene dominantly reflects its molecular electronic structure.

In Fig. 2, we show the valence band spectrum of K-doped picene measured with photon energy of 100 eV. Using incident energy of 100eV was crucial to observe electronic structure change near $E_F$, as the spectral change near $E_F$ was not observed clearly for the spectrum with 1100 eV due to the lower photoionized cross-sections [18]. The highest intensity peak at 19 eV, which overlaps with the C 2*s* dominant bands, is a K 3*p* shallow core level. The spectral shape with a prominent peak around 8 eV and the peak at 14.5 eV is similar to that of pristine picene, though the shape of K-doped picene has less fine structure compared to that of pristine picene. Regarding energy positions, we found that the 8 and 14.5 eV peaks in the K-doped spectrum is larger by 0.5 eV than those in pristine picene. On the other hand, one can see some differences between the two spectra especially in a lower binding energy region ($E_F$-6 eV). As is evident, the steep edge structure around 3.0 eV seen in pristine picene, which corresponds to states derived from the HOMO band, is not clearly observed in the K-doped picene spectrum. This suggests transformation of HOMO band due to K doping. Indeed, we found that the valence band spectrum of K-doped picene, especially for the HOMO derived states ($E_B \approx 3$ eV), is not explained by the calculated spectrum of pristine picene very well. More importantly in the vicinity of $E_F$, a new structure appears in K-doped picene, for which no peak corresponding to this new structure is seen in the spectrum of pristine picene.

Change in the vicinity of $E_F$ by K doping can be seen in Figure 3 in more detail. In pristine picene, a steep spectral cut-off is located at 2.7 eV, corresponds to the HOMO band, and the tail extends to 2.0 eV binding energy, which is referred as the valence band maximum, as shown as a deviation from the zero line in Fig. 3. This binding energy of the valence band maximum is nearly half of the band gap [7], suggesting that solid pristine picene is an intrinsic semiconductor. We also observed no intensity at $E_F$, consistent with semiconducting properties of pristine picene. In K-doped picene, on the other hand, there is a new peak near $E_F$. This result agrees well with the metallic properties of K-doped picene. Note that the new peak near $E_F$ is not a PES signal from K metal resides on the surface region but one intrinsic to K-doped picene, as the spectral shape of the K $3p$ shallow core level (Fig. 2) does not have plasmon features characteristic of K $3p$ shape of K metal. The new peak has a maximum around 0.6 eV and a clear Fermi edge could not be observed within our experimental accuracy. The spectral shape of the new states observed in K-doped picene is reminiscent of the spectra of $K_3C_{60}$ measured at room temperature. The PES spectra of $K_3C_{60}$ show a broad structure around 0.5 eV, which has been attributed to correlation-induced states [9, 19-22]. High-resolution PES studies of $K_3C_{60}$ at lower temperature revealed presence of a metallic edge with fine structures that have been explained with electron-phonon and electron-plasmon interactions [20, 21]. High-resolution PES studies at lower temperature of K-doped picene will provide insight into the superconductivity, as in K-doped fullerides.

Since the energy separation between the bottom edge of the new band near $E_F$ and the valence band maximum of K-doped picene is about 1.0 eV and smaller than the binding energy of the valence band maximum (2.0 eV) of pristine picene, the new peak of K-doped picene can not be explained by simple rigid LUMO band shift of pristine picene due to the filling with electrons. Possibility of doping induced localized states, like in an iodine-doped pentacene [23], as the origin of the new states may be ruled out, since a simple photoionization cross-section analysis showed that the new states have negligible K 4$s$ character. The cross-section of K 4$s$ has $10^{-2}$ times smaller than that of K 3$p$ [11], which would make the intensity of the new states at $E_F$ negligible in the spectrum shown in Fig. 2. In view of the transformation of the HOMO band by K doping, the new peak may be considered as the states derived from the LUMO band of pristine picene changed by K doping. These indicate that HOMO and LUMO bands are modified by K doping, and the electronic structure of K-doped picene is, thus, not obtained by a simple rigid-band shift of that of pristine picene. Non-rigid band like change of band dispersion was predicted from recent first principle calculations [24], where doped-K atoms and more importantly change in the molecular orientation influence the electronic band structure derived from LUMO and LUMO+1 of pristine picene. However, band structure calculations for K-doped picene based on experimental crystal structure data in order to be compared with the present PES spectrum have not been available so far.

For alkali-doped fullerene, importance of electron-intramolecular-vibration interaction has been proposed for observed non-rigid band change, as mentioned above. In order to study the electron-intermolecular vibration interaction in K-doped picene, we discuss the electronic structure of K-doped picene by comparing the overall valence band structure with molecular orbital calculations taking electron-intramolecular-vibration interactions into consideration (Fig. 2). Here we use a calculated result where one electron is doped into a picene molecule (mono-anion picene: $C_{22}H_{14}^-$), as estimated from K concentration $x$ of $1.0 \pm 0.3$. Molecular calculation results of di- and tri-anions (not shown) exhibit over all spectral shape similar to mono-anion but the intensity of the states near $E_F$ depends on the number of electrons. The symmetries of the state near $E_F$ and the next energy level ($E_B \approx 3$ eV) are $a_2$ and $b_1$, respectively, which are the same as the LUMO and HOMO of pristine picene. Electron doping causes changes in the HOMO and LUMO levels in picene, coming from strain of a picene molecule due to a coupling between doped electrons in the LUMO of pristine picene and intramolecular phonon of picene molecule. Thus, the calculated spectrum of electron-doped picene is different from that of pristine picene, evidently at the structure around 3 eV. The observed spectral shape with peaks at 8 and 14.5 eV reveals an almost monotonous decrease of intensity around 2-8 eV, and the structure near $E_F$ are well reproduced with that of the molecular orbital calculation for electron-doped picene as indicated with the vertical solid lines. Excellent agreement between experiments and calculations indicates that electron-intramolecular-vibration coupling between electrons doped into the LUMO and intramolecular phonon of picene molecule makes the electronic structure of K-doped picene different from that of pristine picene. According to the calculations, the electron-phonon coupling constants of picene become stronger with an increase in the number of electrons per picene molecule. This is in line with the observation that K-doped picene ($K_xC_{22}H_{14}$) is found to become a superconductor for $x \approx 3$, corresponding to three electrons doping into picene. These results suggest that the electron-intramolecular-vibration coupling is crucial to understand the physical properties including superconductivity in K-doped picene [25], as discussed for fullerides [10] and BEDT-TTF [26].

## 4. Conclusions

We have performed PES studies of solid pristine and K-doped picene in order to study the electronic structure of solid picene and its evolution with K doping. The valence band spectrum of pristine solid picene consists of mainly three structures that can be ascribed to C $2s$, $2s$-$p$, and $2p$ dominant states, and no states at $E_F$, consistent with semiconducting properties. The spectral shape is well explained with broadened molecular orbital calculations, indicating the molecular nature of solid picene. The valence band spectrum of K-doped picene shows an overall valence band spectral shape similar to that of pristine solid picene with a K 3$p$ shallow core level. However, larger differences are observed for the HOMO derived bands and states near $E_F$, latter of which is consistent with metallic properties. The change of the valence band due to K doping cannot be explained in terms of a simple bandfilling picture with doped electrons. Comparison with available theoretical results shows that the new peak near $E_F$ corresponds to states transformed from LUMO of pristine solid picene due to electron-intramolecular-vibration coupling caused by electron doping. This result points to importance of electron-intramolecular-vibration coupling of the K-doped picene.

**Acknowledgements**

This work was supported partly by a Grant-in-Aid for Scientific Research of the Ministry of Education, Culture, Sports, Science and Technology, Japan (20340091). The measurement at BL27SU of SPring-8 was performed under a proposal number 2009B1697.

[17] This structure can be corresponded to a shoulder structure seen in calculated spectra broadened with Gaussian with a narrower FWHM (not shown). Observable deviation may be related the fact that the orbitals of this region have an in-plane sigma like character and may have larger influence from the induced photoholes. This may be in line with the reduction of this structure in K-doped picene.

[18] PES measurements using 100 eV photon may dominantly reflect a surface derived electronic structure that may be different from bulk electronic structure. As shown in text, however, the observed spectral change due to K doping is consistent with physical properties

of K-doped picene, suggesting that what we observe reflects the bulk electronic structure of K-doped picene.

**Figure captions**

Fig. 1. Incident energy dependence of valence band spectra from pristine picene using photon energy of 1100 and 100 eV, together with molecular orbital calculation of pristine picene. The calculated spectrum was obtained from calculated molecular orbital levels convoluted with Gaussian (FWHM of 0.7 eV) to reproduce the observed spectrum.

Fig. 2. Valence band spectrum from K-doped picene using 100 eV photon energy, together with molecular orbital calculation of mono-anion picene. The calculated spectrum was obtained from calculated molecular orbital levels for mono-anion picene convoluted with Gaussian (FWHM of 0.7 eV) to reproduce the observed spectrum.

Fig. 3. Valence band spectra near $E_F$ from pristine picene and K-doped picene using 100 eV photon energy. The horizontal line of each spectrum denotes a background level.

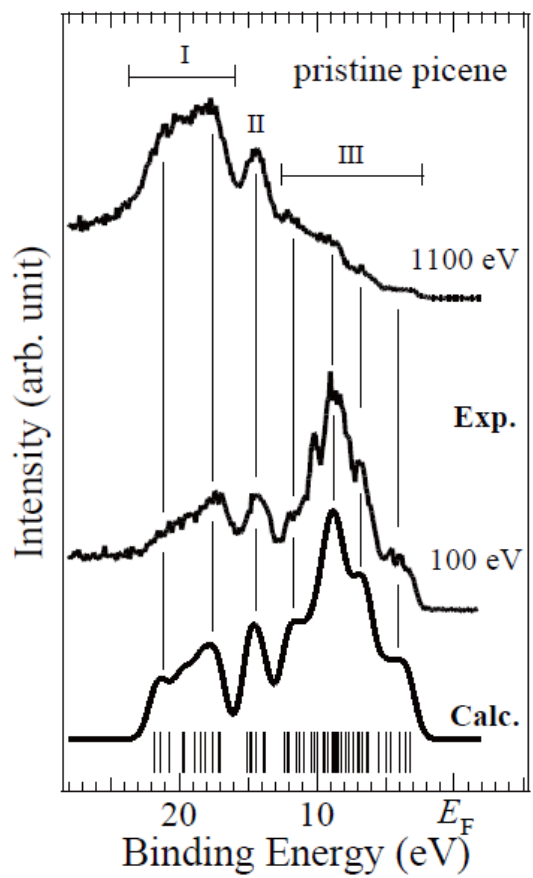

Fig. 1

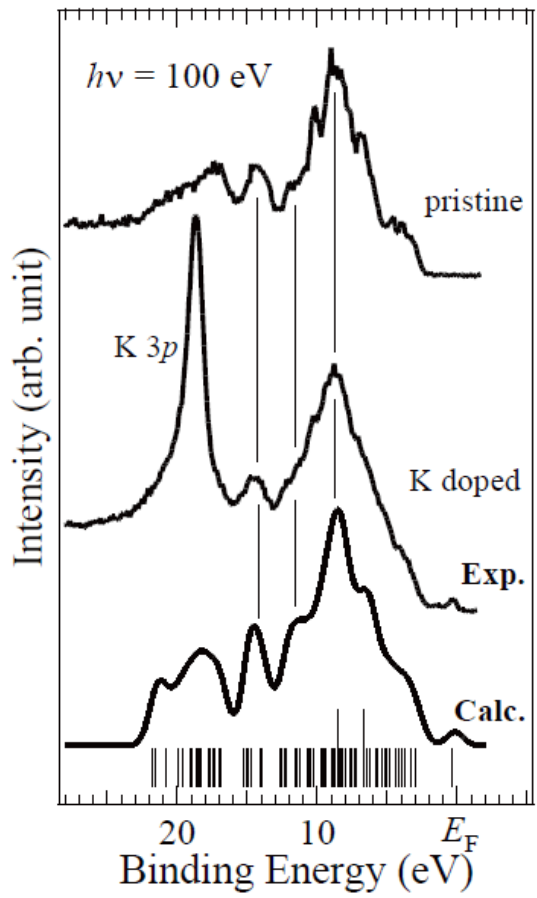

Fig. 2

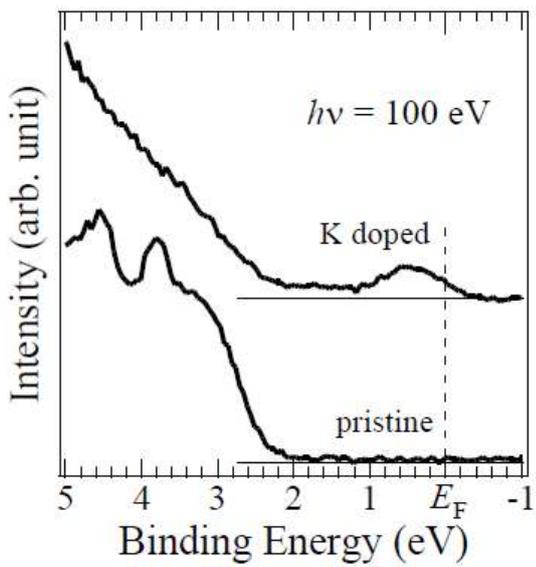

Fig. 3